\documentstyle[11pt]{article}

\begin{document}
\begin{center}

{\LARGE The proton spin and the Wigner rotation}

\vspace{15mm}

\renewcommand{\thefootnote}{\fnsymbol{footnote}}
{\large Bo-Qiang Ma\footnote
{Fellow of Alexander von Humboldt Foundation.}$^{,1}$
and Qi-Ren Zhang$^{2}$}

\vspace{5mm}

\vspace{5mm}
{\small $^{1}$Institute of High Energy Physics,
Academia Sinica, P.O.Box 918(4),
Beijing 100039, China and Institut f\"ur Theoretische Physik der
        Universit\"at Frankfurt am Main, Postfach 11 19 32,
        D-6000 Frankfurt, Germany}

\vspace{5mm}
{\small $^{2}$Center of Theoretical Physics, CCAST (World Laboratory),
Beijing, China and Department of Technical Physics, Peking University,
Beijing 100871, China }

\vspace{14mm}

\end{center}

\begin{minipage}[t] {105mm}
  {\large \bf Abstract } \\

\vspace{4mm}
It is shown that in both the gluonic
and strange sea explanations of the Ellis-Jaffe sum rule
violation discovered by the European Muon Collaboration (EMC), the
spin of the proton, when viewed in its rest reference frame, could
be fully provided by quarks and antiquarks within a simple quark model
picture,
taken into account the relativistic effect from the
Wigner rotation.
\end{minipage}

\vspace{20mm}
{To be published in Z.~Phys.~C}

\break

The European Muon Collaboration (EMC) experiment \cite{EMC} on deep
inelastic scattering of polarized leptons from polarized protons has
received attention by the particle physics community
recently\cite{Reya}. The
reason is that the result of the integrated
spin-dependent structure function data is
significantly smaller than that expected from the Ellis-Jaffe sum
rule\cite{EJ,Jaf87},
and that in a naive interpretation of this small result
one is led to the conclusion \cite{EMC,Lea88}
that the fraction of the proton spin
carried by quarks and antiquarks is smaller than expected.
There are essentially
two theoretical aspects for understanding this startling EMC result.
The first concerns the mechanism for the violation of the Ellis-Jaffe
sum rule (EJSR), which is obtained \cite{EJ,Jaf87} by using the
experimental values of $G_{A}/G_{V}$ measured in weak semileptonic
decays together with flavor SU(3) symmetry results
and an additional assumption of a negligible strange sea
contribution
to the proton spin. Many theoretical speculations have been proposed
to explain the EJSR violation, such as to attribute the small EMC
result to gluonic contributions
due to the U(1) axial anomaly\cite{Jaf87,gluon},
non-vanishing strange quark contributions\cite{strange},
consequences of the QCD anomaly and the spontaneous breaking of
the chiral symmetry \cite{Fri},
the unjustified $x\rightarrow0$ extrapolation of
the data\cite{smallx},
flavor SU(3) symmetry
breaking\cite{su3b}, and non-perturbative QCD effects at
low $Q^{2}$\cite{Jaf87,Kun89,Pre88} et al..
This paper is
not intended to discuss the aspect concerning the EJSR violation.
Our
attention is focused on the second issue concerning the EMC result:
the small data triggered the proton ``spin crisis''\cite{Lea88},
i.e., the
intriguing question of how the spin of the proton is distributed
among
its quark spin, gluon spin and orbital angular momentum. At present
it is commonly taken for granted that the EMC result implies that
there must be some contribution due to gluon polarization or orbital
angular momentum to the proton spin. For example, in the gluonic
\cite{gluon}
and strange sea \cite{strange}
explanations of the EJSR breaking, the proton
spin carried by the spin of quarks and antiquarks
was estimated to be of about
$70\%$
in the former \cite{Tor89} and negligible in the
latter\cite{strange}. We will show,
however, that the above understandings are not in contradiction
with a simple quark model picture in which the spin of the proton,
when viewed in its
rest reference frame, is fully provided by the vector sum of the
spin of quarks and antiquarks.

The key points for understanding the proton spin puzzle lie in the
facts that the vector sum of the constituent  spin for a composite
system is not Lorentz invariant by taking into account the
relativistic effect from the Wigner rotation\cite{Wigner},
and that it is in the
infinite momentum frame the small EMC result was interpreted
\cite{EMC,strange} as
an indication that quarks and antiquarks
carry a small amount of the total spin of
the proton. From the first fact we know that the vector spin
structure of
hadrons could be quite different in different frames from the
relativistic viewpoint. We thus can understand the proton
``spin crisis'', because there is no need to require that the sum of
the spin of quarks and antiquarks
be equal to the proton spin in the infinite
momentum frame, even if the vector sum of the spin  of quarks and
antiquarks equals to the
proton spin in the rest frame. In fact this idea has already been
presented \cite{Ma} in a light-cone language in which the effect of
the Wigner
rotation manifests itself as the effect of the Melosh
rotation\cite{Melosh}. It
was shown that the small EMC result could be naturally reproduced
within the SU(6) naive quark model by taking into account the effect
of the Melosh rotation without contributions due to gluon spin and
orbital angular momentum.  The validity of the Bjorken sum
rule [17] may be also retained if one imposes some further
constraints on the flavor distribution of quarks in the proton.
However, the work of Ref.\cite{Ma} consider neither the weak hyperon
decay data used in previous analyses nor the mechanism
that breaks the EJSR. The analysis was also presented in a
particular (though most convenient) language which might be
considered
as a model-dependent framework with no universal
significance. Therefore we need to re-analyze this issue in
conventional languages.

We now explain why the vector sum of the constituent  spin for a
composite
system is not Lorentz invariant. It should be kept in mind that the
notion of spin is essentially a relativistic notion associated with
the space-time symmetry group of Poincar\'e\cite{Wigner}.
In the relativistic
dynamical theory of the quantum system\cite{Dirac},
measurable physical
quantities are closely related to the ten generators of the
Poincar\'e group. The explicit representation of generators
relies on the form of dynamics\cite{Dirac,Leu78}.
In the conventional
instant form dynamics, the generators $P^{\mu}=(H,\vec{P})$ of the
space-time translations have the physical significance of energy and
momentum, and the generators $J^{\mu\nu}$ of the infinitesimal
Lorentz transformations are related to the angular momentum
$\vec{J}$ by $J^{k}=(1/2)\epsilon_{ijk}J^{ij}$ and the boost
generator $\vec{K}$ by $K^{k}=J^{k0}$. The conventional
3-vector spin $\vec{s}$ of a moving particle with finite mass
$m$ and
4-momentum $p_{\mu}$ can be defined by transforming its
Pauli-Lub\'anski 4-vector
$w_{\mu}=(1/2)J^{\rho\sigma}p^{\nu}\epsilon_{\nu\rho\sigma\mu}$
\cite{Lub42}
to its rest frame via a rotationless Lorentz boost $L(p)$, which
satisfies $L(p)p=(m,\vec{0})$, by $(0,\vec{s})=L(p)w/m$
\cite{Coe82}.
Under an
arbitrary Lorentz transformation, a particle of spin $\vec{s}$ and
4-momentum $p_{\mu}$ will transform to the state of spin
$\vec{s'}$ and 4-momentum $p'_{\mu}$ by
\begin{equation}
  \vec{s'}=\Re_{w}(\Lambda,p)\vec{s},\;\;\;\;\;p'=\Lambda\:p,
\end{equation}
where $\Re_{w}(\Lambda,p)=L(p')\Lambda\:L^{-1}(p)$ is a pure rotation
known as the Wigner rotation. For simplicity we assume that the
proton is
a composite system of moving quarks and antiquarks
and that the proton spin is fully
provided by the spin of quarks and antiquarks
in the proton rest frame. When the proton
is boosted, via a rotationless Lorentz transformation along its spin
direction, from the rest frame to a frame where the proton is
moving, each quark spin will undergo a Wigner rotation, and these
spin rotations may produce, arising from the relativistic effect due
to internal quark motions, a significant change in
the vector sum of the
spin of quarks and antiquarks.
The proton spin, however, remains the same as that in the
rest frame according to the spin definition. In consequence the
vector sum
of the spin of quarks and antiquarks
may differ non-trivially from the proton spin in the
new frame.

It then becomes necessary to clarify what is meant by the quantity
$\Delta q$ defined by
$\Delta q\!\cdot\!S_{\mu}=<\!P,S|\bar{q}
\gamma_{\mu}\gamma_{5}q|P,S\!>$,
where
$S_{\mu}$ is the proton polarization vector. $\Delta q$ can be
calculated from $\Delta
q=<\!P,S|\bar{q}\gamma^{+}\gamma_{5}q|P,S\!>$,
as the instantaneous fermion lines do not contribute to the +
component\cite{Ma91}. One can easily prove,
by expressing the quark wave
functions in terms of light-cone
Dirac spinors\cite{LC} (i.e., the quark
spin states in the infinite momentum frame), that
\begin{equation}
\Delta q=\int_{0}^{1}dx\:[q^{\uparrow}(x)-q^{\downarrow}(x)],
\end{equation}
where $q^{\uparrow}(x)$ and $q^{\downarrow}(x)$ are probabilities
of finding, in the proton infinite momentum frame, a quark or
antiquark of flavor q with fraction x of the proton longitudinal
momentum and with polarization parallel
and antiparallel to the proton
spin, respectively. However, if one expresses the quark wave
functions in terms of conventional instant form Dirac spinors
(i.e., the quark spin states in the proton rest frame, with the
normalization condition $\bar{u}\gamma^{+}u=-\bar{v}\gamma^{+}v=1$),
it can be
found, that
\begin{equation}
\Delta q=\int\!d^{3}\vec{p}
\:M_{q}\:[q^{\uparrow}(p)-q^{\downarrow}(p)]
=<\!M_{q}\!>\Delta q_{L},
\end{equation}
with
\begin{equation}
M_{q}=[(p_{0}+p_{3}+m)^{2}-\vec{p}_{\perp}^{2}]/[2(p_{0}+p_{3})
(m+p_{0})]
\end{equation}
being the contribution from the relativistic effect due to quark
transversal motions, $q^{\uparrow}(p)$ and $q^{\downarrow}(p)$ being
probabilities of finding, in the proton rest frame, a quark or
antiquark of flavor $q$ with rest mass $m$ and momentum $p_{\mu}$ and
with spin parallel and antiparallel to the proton spin
respectively, and $\Delta q_{L}=\int\!d^{3}\vec{p}
[q^{\uparrow}(p)-q^{\downarrow}(p)]$ being the net spin vector sum of
quark flavor $q$ parallel to the proton spin in the rest frame. Thus
one sees that the quantity $\Delta q$ should be interpreted as the
net spin polarization in the infinite momentum frame if one properly
considers the relativistic effect due to internal quark transversal
motions.

{}From the above considerations, one naturally reaches the
conclusion that the spin structure of a composite system should be
defined in the rest frame of the
system with internal constituent motions also taken into account.
Thereby we can understand the ``spin crisis'',
simply because the quantity
$\Delta\Sigma=\Delta u+\Delta d+\Delta s$ represents, in a strict
sense, the sum of the quark helicity
(or longitudinal-component spin) in the
infinite momentum frame rather than the vector
sum of the spin carried
by quarks and antiquarks in the proton rest frame.
It is possible that the value of
$\Delta\Sigma=\Delta u+\Delta d+\Delta s$ is small whereas the
simple spin sum rule
\begin{equation}
\Delta u_{L}+\Delta d_{L}+\Delta s_{L}=1
\end{equation}
still holds. By eq.(5) we mean a simple quark model picture in which
the proton spin is fully provided by quarks and antiquarks in the
proton rest frame.
The gluon spin and the orbital angular momentum may
contribute to the proton spin in the rest frame.  However, one
may reasonably speculate that the proton, being the most stable
hadron in the real world, can be described in
the simple quark model picture with the vector sum of the spin of
quarks and antiquark
being the proton spin.
We will show that this philosophy could be apply to both the
gluonic \cite{gluon} and strange sea \cite{strange}
explanations of the EJSR
breaking, though the real situation might be complicated.

Theoretically, the integrated spin-dependent structure function of
the proton is related through short-distance expansion to the flavor
isotriplet, octet and singlet components of the axial vector matrix
elements, $\Delta q^{3}$, $\Delta q^{8}$ and $\Delta q^{0}$, by
the relation[2,3]
\begin{equation}
\int_{0}^{1}dx\:g_{1}^{p}(x)=(1/12)\Delta q^{3}+(1/36)\Delta q^{8}
+(1/9)\Delta q^{0}.
\end{equation}
The two non-singlet axial charges, $\Delta q^{3}$ and
$\Delta q^{8}$, can be inferred by \cite{BSR}
\begin{equation}
\Delta q^{3}=\Delta u-\Delta d=G_{A}/G_{V}=1.261
\end{equation}
from neutron decay \cite{Data} plus isospin symmetry, and by
\begin{equation}
\Delta q^{8}=\Delta u+\Delta d-2\Delta s=0.675
\end{equation}
from strangeness-changing hyperon decays \cite{Bou83} plus flavor SU(3)
symmetry. Prior to the EMC experiment, the flavor singlet (or the
U(1)) axial charge was evaluated by Ellis-Jaffe and Gourdin\cite{EJ},
suggesting
$\Delta s=0$, to be
$\Delta q^{0}=\Delta\Sigma=\Delta u+\Delta d+\Delta s
=\Delta q^{8}$. Then one obtains, neglecting small QCD corrections,
the EJSR
$\int_{0}^{1}dx\:g_{1}^{p}(x)=0.198$, a value which is significantly
larger than the revised EMC result
$\int_{0}^{1}dx\:g_{1}^{p}(x)=0.126$.

A possible explanation of this discrepancy was proposed \cite{gluon}
by adding
to the ``naive'' U(1) charge,
$\Delta\Sigma=\Delta u+\Delta d+\Delta s$, a gluonic contribution due
to the Adler-Bell-Jackiw anomaly, $\Delta
q^{0}=\Delta\Sigma-(\alpha_{s}/2\pi)n_{f}\Delta g$, where $\Delta
q^{0}$ is the physical U(1) charge used in the Ellis-Jaffe relation
eq.(6), $\alpha_{s}$ is the QCD coupling constant, and $n_{f}$ is the
number of excited flavors. It is expected that $\Delta\Sigma\approx
\Delta q^{8}$ and that the gluonic contribution,
$(\alpha_{s}/2\pi)n_{f}\Delta g$, nearly compensates
$\Delta\Sigma$. Then one obtains, combining eqs.(7),(8) and eq.(6)
inferred by the EMC result, that
\begin{equation}
\Delta u=0.968,\;\;\;\Delta d=-0.293,\;\;\;\Delta s=0,
\end{equation}
and
\begin{equation}
\Delta\Sigma=\Delta u+\Delta d+\Delta s=0.675.
\end{equation}
This result was interpreted \cite{gluon,Tor89}
as an implication that a large
fraction ($\Delta\Sigma\approx 70\%$) of the proton spin is carried by
the spin of quarks and antiquarks balanced by a sizable contribution
($1-\Delta\Sigma\approx 30\%$) arising from the gluon spin and/or from
orbital angular momentum. An alternative explanation of the small EMC
result was proposed \cite{strange}, based on the Skyrme model, by still
attributing $\Delta q^{0}=\Delta\Sigma$ but suggesting a large
$\Delta s$. Combining eqs.(7), (8) and eq.(6) inferred by the EMC
result, one obtains
\begin{equation}
\Delta u=0.750,\;\;\;\Delta d=-0.511,\;\;\;\Delta s=-0.218,
\end{equation}
and
\begin{equation}
\Delta\Sigma=\Delta u+\Delta d+\Delta s=0.020.
\end{equation}
This result was interpreted \cite{strange}
as an implication that most of the
proton spin is due to gluons and/or orbital angular momentum. These
interpretations of the proton spin structure triggered some further
studies \cite{Lea88,Dzi89}
about how the proton spin is distributed among the
spin and orbital angular momentum of its quarks, antiquarks and gluons.

We now show that in both of the above explanations of the
EJSR violation
the spin of the proton could be fully provided by the vector sum of
the spin of quarks and antiquarks in the simple quark model picture
based on our above
definition of the spin structure of a composite system and above
clarification of the physical implication of the quantity $\Delta q$.
Since $<\!\!M_{q}\!\!>$, the average contribution from
the relativistic
effect due to internal transversal motions of quark flavor $q$,
ranges from $0$ to $1$, and $\Delta q_{L}$,
the net spin vector polarization
of quark flavor $q$ parallel to the proton spin in the proton rest
frame, is related to the quantity $\Delta q$ by the relation
$\Delta q_{L}=\Delta q/<\!\!M_{q}\!\!>$,
we have sufficient freedom to make
the simple quark model spin sum rule,
i.e., $\Delta u_{L}+\Delta d_{L}+\Delta s_{L}=1$,
satisfied while still preserving the values of $\Delta u$,
$\Delta d$ and $\Delta s$, i.e., eqs.(9) and (11), in the two
explanations, respectively. In the gluonic explanation, we could
choose $\Delta u_{L}=4/3$, $\Delta d_{L}=-1/3$, and $\Delta s_{L}=0$
as those in the most simply SU(6) configuration
of the naive quark model,
then we need
$<\!\!M_{u}\!\!>\approx <\!\!M_{d}\!\!> \approx 0.7$ to
preserve eq.(9). It can be found that we could reproduce such
$<\!\!M_{u}\!\!>$ and $<\!\!M_{d}\!\!>$ by using a quark mass $m
\approx 330$ MeV and $\sqrt{\vec{p}_{\perp}^{2}} \approx 330$ MeV
in the constituent quark model framework.
In the strange sea explanation, the non-vanishing
$\Delta s$ reflects polarizations of sea thus some number of
sea
quarks should be introduced.
Then one has large freedom to choose arbitrary
$\Delta u_{L}$, $\Delta d_{L}$, and $\Delta s_{L}$ constrained by the
simple quark model spin sum rule eq.(5)
while still preserving eq.(11). In both of the
above cases the proton spin is fully provided by the vector sum of
the spin of quarks and antiquarks.

We need to clarify the seeming contradiction
between the statements in
this paper and those in much of the literature about the fraction of
the proton spin carried by quarks and antiquarks.
The difference is arising from the
definitions of spin. In this paper we refer the spin of a moving
quark to the conventional 3-vector spin defined by transforming its
Pauli-Lub\'anski 4-vector to its rest frame. Thus a quark or
antiquark, when simply described by
a conventional instant form Dirac
spinor (i.e., $u^{\uparrow,\downarrow}(p)$ for quark and
$v^{\uparrow,\downarrow}(p)$ for antiquark),
provides 1/2 net spin contribution
parallel or antiparallel to the proton spin. Whereas in the
literature the spin operator of quarks is referred to
$\bar{\psi}\gamma_{\mu}\gamma_{5}\psi$, which is essentially the
non-conserved axial vector current. From this
definition a quark state expressed by
$u^{\uparrow,\downarrow}(p)$ or
$v^{\uparrow,\downarrow}(p)$
contributes only a value of $(1/2)M_{q}$ net spin
contribution parallel or antiparallel to the proton spin. The
reason for this reduction of spin contribution
can be ascribed to a negative spin contribution
from the lower component of the Dirac spinor if the
quark transversal motions are
considered.  However, the lower component is considered to have
again a positive orbital angular momentum in this situation, and in
sum the total contribution
(i.e., spin+orbital angular momentum) from
this quark state to the proton spin is the same as that of
the 3-vector spin defined in our paper.  Thus the ``spin crisis''
could be understood within the simple quark model picture by
considering the relativistic effect from internal quark
transversal motions.

We have made many simplifications
in the above discussions of the proton spin problem.
The most important
simplification is that we treat the Wigner rotation in a free quark
approximation and do not consider any dynamical effect due to quark
interactions in the boost from the proton rest frame to the
infinite momentum frame.
However, the inclusion of dynamical effects  can
only change the results (e.g., the explicit expression
of $M_{q}$) quantitatively. It should not affect the qualitative
conclusions, because the effects from kinematics
should be first considered before
the introduction of other dynamical effects.
We are still far from the
answer to how the proton spin is distributed among the spin and
orbital angular momentum of quarks, antiquarks and gluons.
We only provide two
toy cases in a simple quark model picture in which the proton spin is
fully provided  by quarks and antiquarks with
the EMC result of the proton spin-dependent structure function also
satisfied. For further understanding of the proton spin structure we
need more theoretical and experimental works.

In summary, we discussed the Wigner rotation in the
spin structure of a composite system. We showed that the proton spin
puzzle caused by the EMC polarized muon proton data could be
understood within a simple quark model picture in
which the proton spin,
when viewed in its rest reference frame, is fully provided by
quarks and antiquarks, taking into account the relativistic
effect due to internal quark transversal motions.

\newpage

\noindent
{\bf Acknowledgment}

We are grateful to Prof.W.Greiner for his hospitality and for the
support from Institut f\"ur Theoretische Physik der
Universit\"at Frankfurt, where this work is completed. We thank
Prof.A.Sch\"afer and Dr.L.Mankiewcz for the discussions.
One of the author (B.-Q.M) would like to thank the very helpful and
inspiring discussions
with Prof.Tao Huang and many colleagues in Beijing, where this work
was started.
He also thanks the hospitality of Prof.F.P\"uhlhofer during
his stay in Marburg.  We thank the referee for his encouragements
and valuable comments.

\break

\end{document}